\documentclass[%
 reprint,
 amsmath,amssymb,
 aps,
 prl,
 floatfix,
]{revtex4-1}

\usepackage{graphicx}
\usepackage{dcolumn}
\usepackage{bm} 
\newcommand{\vcr}[1]{\bm{\mathrm {#1}}} % vector notation
\newcommand{\breathe}[2]{\rule[-#1ex]{0cm}{#2ex}}
\newcommand{\sst}[2]{{#1}_{\text{#2}}}
\newcommand{\argmin}[1]{\stackrel[#1]{}{\text{argmin}}}

\begin{document}

\preprint{APS/123-QED}

\title{Machine Learning Regularized Solution of the Lippmann-Schwinger Equation}

\author{Subeen Pang}
\email{sbpang@mit.edu}
\author{George Barbastathis}
\altaffiliation[Also at: ]{Singapore-MIT Alliance for Research and Technology (SMART) Centre, 1 CREATE Way, Singapore 138602, Singapore.}
\affiliation{Mechanical Engineering, Massachusetts Institute of Technology, Cambridge, Massachusetts 02139, USA}

\date{\today}

\begin{abstract}
Solution of the discretized Lippmann-Schwinger equation in the spatial frequency domain involves the inversion of a linear operator specified by the scattering potential. To regularize this inevitably  ill-conditioned problem, we propose a machine learning approach: a Recurrent Neural Network with long short-term memory (LSTM) and with the null space projection of the Lippmann-Schwinger kernel on the recurrence path. The learning method is trained using examples of typical scattering potentials and their corresponding scattered fields. We test the proposed method in two cases: electromagnetic scattering by dielectric objects, and electron scattering by multiple screened Coulomb potentials. In both cases the solutions to test examples, disjoint from the training set, were obtained with fewer iterations and were accurate compared to linear solvers. We also observed surprising generalization ability: in the electromagnetic case, an LSTM trained with random arrangements of dielectric spheres was able to obtain the correct solutions for general topologically similar objects, such as polygons. This suggests that the LSTM successfully incorporates the physics of scattering into the inversion algorithm.
\end{abstract}

\maketitle

The Fredholm integral equation of the second kind, in physics more commonly known as Lippmann-Schwinger equation (LSE)
\begin{equation}\label{eq:LippSch}
    \psi(\vcr{r})=\sst{\psi}{i}(\vcr{r})+
    \int G\left(\vcr{r}-\vcr{r'}\right)V\left(\vcr{r'}\right)
    \psi\left(\vcr{r'}\right)\:\text{d}\vcr{r'},
\end{equation}  
results from several classical and quantum scattering principles. It relates the scattered field $\psi$ to the incident field $\sst{\psi}{i}$, scattering potential $V$ and free-space Green's function $G$. After adequate spatial sampling of $\psi$ to a discrete wavefunction $x$, Eq.~(\ref{eq:LippSch}) may be converted to a linear inverse problem \cite{vainikko2000fast,zepeda2016fast,ying2015sparsifying} 
\begin{equation} \label{eq:Ax=b}
    Ax=b.
\end{equation}
Similar linear problems result from numerous partial differential equations and integral equations\ \cite{gockenbach2005partial, solin2005partial} and extensive research has been devoted to linear system solvers\ \cite{eason1976review,gould2007numerical}. 

Since the late 1990s, the solution of linear inverse problems has been increasingly informed by the branch of function approximation theory that deals with sparse representations \cite{starck2009overview}. Sparsification of the wavefunction sometimes is appropriately {\it ad hoc}, {\it e.g.} using redundant bases such as wavelets \cite{mallat1999wavelet}; whereas at other times it is best learnt from data. This last approach requires several examples of typical solutions $x$ and their corresponding ``measurements'' $b$ to be available, and the resulting algorithms are referred to as dictionaries \cite{aharon2006k,inv:elad2006image}. The inverse estimate is obtained as 
\begin{equation} \label{eq:regoptimproblem}
    \hat{x} = \:\argmin{x} \left\{\breathe{0}{2.5} \left\lVert Ax - b \right\rVert^2 + \kappa \Phi(x)\right\},
\end{equation}
where $\Phi(x)$ is the sparsity-promoting regularizer, and $\kappa$ is the regularization parameter. Typical such regularizers include the $L_1$ norm, {\it i.e.} are non-differentiable. The solution of Eq.~(\ref{eq:regoptimproblem}) proceeds iteratively in computational (discrete) time $t$ according to the proximal gradient method \cite{inv:daubechies04} as 
\begin{equation} \label{eq:proxalg}
    \hat{x}[{t+1}] = \mathcal{P}_{\Phi} \left\{ \breathe{0}{2.5} \alpha A^{\dagger}b + \left( I-\alpha A^{\dagger}A \right)\hat{x}[{t}] \right\}
\end{equation}
or one of its speedier relatives such as FISTA \cite{beck2009fast}. The proximal operator $\mathcal{P}_{\Phi}$ is derived from the regularizer $\Phi.$ The second term in the bracket is the null space projection, and it can be thought of as encoding the physical model $A$ into the learning scheme.

\begin{figure}[t!]
    \centering
    \includegraphics{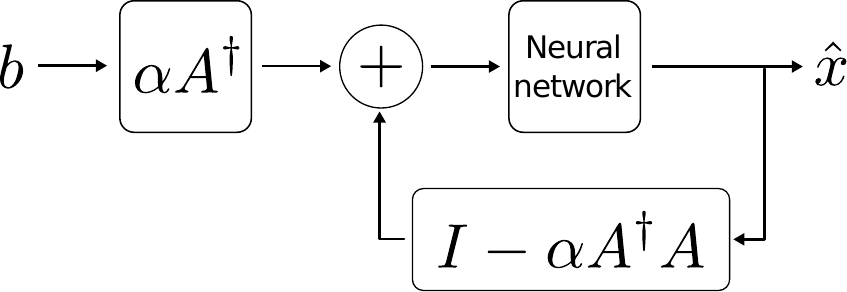}
    \caption{Neural network regularized proximal gradient method.}
    \label{fig:nnprox}
\end{figure}

In 2016, it was proposed that, instead of using the $L_1$ norm regularizer $\kappa$ chosen in an {\it ad hoc} manner and then running Eq.~(\ref{eq:proxalg}), sparsifiers even better fit to available data $\left\{x,b\right\}$ may be obtained by learning $\mathcal{P}_{\Phi}$ which is represented by a linear combination of basis functions  \cite{kamilov2016learning}. Extending the space of functions that can be learnt, a neural network is used to approximate $\mathcal{P}_{\Phi}$ \cite{mardani2017recurrent}. This is schematically represented in Fig.~\ref{fig:nnprox}. However, implementations of this idea to actual linear inverse problems Eq.~(\ref{eq:Ax=b}) soon realized that the recurrence destabilizes the neural network's training \cite{hochreiter1991untersuchungen}. Thus, they resorted to breaking the recurrence into a cascade, occasionally trained in adversarial fashion \cite{mardani2017recurrent}. The cascade means that now multiple neural networks need to be trained; and the number of recurrences allowed needs to be built-in the architecture irrevocably. Here, we acknowledge that recurrent dynamics, as in Fig.~\ref{fig:nnprox}, are stabilized if one replaces the neural network with a long short-term memory (LSTM) $\mathcal{C}_{\text{LSTM}}$ whose key function is selective temporal forgetting \cite{hochreiter1997long}. 

Thus, to solve (\ref{eq:Ax=b}) we apply the recurrence
\begin{equation} \label{eq:RNNproximal}
    \hat{x}[{t+1}] = \mathcal{C}_{\text{LSTM}} \left\{ \breathe{0}{2.5} \alpha A^{\dagger}b + \left( I-\alpha A^{\dagger}A \right)\hat{x}[{t}] \right\}.
\end{equation}
We find that, applied to scattering problems, the recurrent method Eq.~(\ref{eq:RNNproximal}) provides accurate solutions while typically requiring fewer iterations than linear solvers such as the generalized minimal residual method and the conjugate gradient method (CG); and generalizes surprisingly well to scattering potentials significantly deviating from the training set. We attribute this latter property to the physical operator $A$ becoming strongly ingrained into the LSTM through the recurrent learning process. The approach of Eq.~(\ref{eq:RNNproximal}) for solving (\ref{eq:Ax=b}) may be applicable for more general physical systems; we chose here the LSE, as an especially acute case that becomes ill-conditioned as the frequency increases and when multiple scattering cannot be neglected \cite{zepeda2016fast,ying2015sparsifying}.

We perform a numerical experiment to test the applicability of Eq.~(\ref{eq:RNNproximal}) to solve physical problems of type Eq.~(\ref{eq:Ax=b}). As an example, the scattering of the incident wave $\sst{\psi}{i}(\vcr{r})=e^{ikz}$, where $\lambda=2\pi/k$ is the wavelength, is studied. We consider two types of the potential $V$. First,
\begin{equation}
    V(\vcr{r}) = k^2\left[n^2(\vcr{r})-1\right],
\end{equation}
which describes the optical scattering in isotropic and linear media with refractive index $n$; and second,
\begin{equation}
    V(\vcr{r}) = 2\sum_{j=1}^{P} \frac{\operatorname{erf}(\eta|\vcr{r}-\vcr{r}_j|)}{|\vcr{r}-\vcr{r}_j|}
    e^{-\xi |\vcr{r}-\vcr{r}_j|}
    \label{eq:LSCoulomb}
\end{equation}
which represents screened Coulomb potential on an electron from $P$ unit charges with a screening constant $\xi$. The error function $\operatorname{erf}$ with a parameter $\eta$ is used to regularize the singularities at $\vcr{r}_j$ without severely compromising physical accuracy \cite{gonzalez2016smooth}. For the numerical evaluation of the LSE, we consider a cubic domain $\Omega=\left\{ \vcr{r}=(x,y,z) \middle| -L/2 \leq x,y,z < L/2 \right\}$ with $N^3$ voxels. On $\Omega$, the LSE is discretized as Eq.~(\ref{eq:Ax=b}) by expanding $\psi$ and $\sst{\psi}{i}$ in the Fourier basis and substituting the integral with the discrete convolution \cite{vainikko2000fast, pham2020three}. More information on the numerical implementation is provided in the Supplemental Material. Eq.~(\ref{eq:Ax=b}) is iteratively solved with the CG until the relative $L_2$-norm error $\left\|\psi-\sst{\psi}{i} \right\|_2/\left\| \sst{\psi}{i} \right\|_2$ reaches $10^{-6}$.

$\mathcal{C}_{\text{LSTM}}$ is trained for each type of potential. 9600 pairs of $(\psi, V)$ are generated by the CG and split into 9000 training, 300 validation, and 300 test examples. For the optical scattering problem, $k$, $L$, and $N$ are $2\pi/\lambda$, $12$, and $64$, respectively where $\lambda$, the wavelength, is 0.532. The refractive index $n$ consists of 1 to 3 spherical objects whose magnitude ranges from 1.01 to 1.03 and diameter from 4 to 8. For Coulomb scattering problem, $P$ ranges from 1 to 3. $k$, $N$, $L$, $\eta$ and $\xi$ are 9, 64, 12, 15 and 0.4, respectively. The training loss function is the negative Pearson correlation coefficient \cite{goy2019high}. Starting from $\hat{x}[{0}]=\sst{\psi}{i}$, the recurrence in Eq.~(\ref{eq:RNNproximal}) continues for four steps, {\it i.e.} until $\breathe{0}{2.5} \alpha A^{\dagger}b + \left( I-\alpha A^{\dagger}A \right)\hat{x}[{3}]$. The detailed implementation of $\mathcal{C}_{\text{LSTM}}$ is shown in the Supplemental Material. To assess the quality of the different objects $\theta$ and $\phi$, we show three different metrics: the relative error defined as $\epsilon(\theta,\phi)=\frac{1}{J}\sum_{j=1}^{J}\left\lvert \frac{ \theta(j)-\phi(j) }{\operatorname{max}[\phi]}  \right\rvert$, the structural similarity index (SSIM) \cite{wang2004image}, and the peak signal-to-noise ratio (PSNR). We assess not only the phase and the amplitude of complex fields but the phase unwrapped along the optical axis ($z$) to prevent any artifacts arising from the phase discontinuity.

\begin{table}[t!]
    \caption{\label{tab:optical_lstm} The image quality assessment of the LSTM on the 300 test examples for the optical scattering problem. The scattered fields estimated by solving the LSE with the CG are set as the ground truth images. Relative error in this table means the relative error per voxel.}
    \begin{ruledtabular}
    \begin{tabular}{cccc} % left, center, decimal, right
    Component & Relative error & SSIM & PSNR \\
    \colrule
    Amplitude & 0.008 & 0.940 & 38.525 \\
    Phase & 0.006 & 0.9915 & 25.646 \\
    Unwrapped phase & $9.3\times 10^{-5}$ & 1.000 & 77.554 \\
\end{tabular}
\end{ruledtabular}
\end{table}

\begin{table}[t!]
    \caption{\label{tab:cmp_shp} The image quality assessment of the LSTM on the 6 objects consisting of polyhedral refractive index profiles for the optical scattering problem. The detalied information on the shapes of these objects is in the Supplemental Material. The scattered fields estimated by solving the LSE with the CG are set as the ground truth images. Relative error in this table means the relative error per voxel.}
    \begin{ruledtabular}
    \begin{tabular}{cccc} % left, center, decimal, right
    Component & Relative error & SSIM & PSNR \\
    \colrule
    Amplitude & 0.008 & 0.920 & 38.671\\
    Phase & 0.006 & 0.989 & 24.841\\
    Unwrapped phase & $1.0\times 10^{-4}$ & 1.000 & 77.179\\
    \end{tabular}
    \end{ruledtabular}
\end{table}

\begin{figure*}[tbp!]
    \centering
    \includegraphics[width=\textwidth]{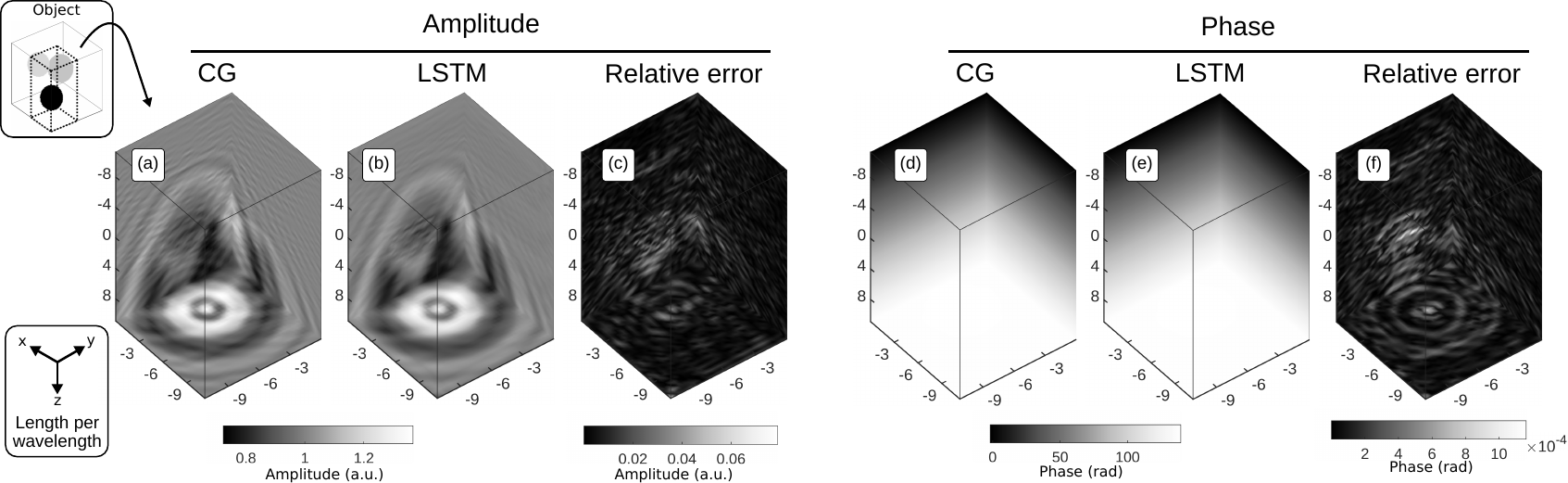}
    \caption{Comparison of the LSE with the CG and the LSTM on a test object consisting of three spheres for the optical scattering problem. The computational box and the location of the object is drawn in the subfigure at the left top corner. The scattered field in the dotted region of this subfigure is shown here. (a),(b) The CG and the LSTM estimations on the amplitude. (c) The relative error in the amplitude. (d),(e) The CG and the LSTM estimations on the unwrapped phase. (f) The relative error in the unwrapped phase. An unit length in this figure corresponds to $\lambda$.}
    \label{fig:compareLSTM_spheres}
\end{figure*}

\textit{Numerical study on optical scattering.}---To test the validity of the CG solutions, they are compared to the Mie theory and the finite-difference time-domain method in the Supplemental Material. In overall, the LSE solved with the CG can well estimate the Mie and finite-difference time-domain solutions with tolerable errors.

The scattered fields generated by the LSTM are shown in Fig.~\ref{fig:compareLSTM_spheres}. As can be seen in Figs.~\ref{fig:compareLSTM_spheres}(a)-(c), the LSTM results well approximate the CG solutions with a slight loss of high-frequency features in the amplitude, while retaining major interference patterns. Table \ref{tab:optical_lstm} summarizes the image quality metrics for the 300 test samples, which shows that the relative error per voxel is less than 0.01 for both the amplitude and the phase. Achieving such low error implies that the LSTM has successfully learnt the appropriate regularization for the spherical objects. In the Supplemental Material, the LSTM results are compared with those from the U-Net \cite{ronneberger2015u}, one of the state-of-art neural network architectures used for image segmentation and image regression tasks \cite{guan2019fully, deng2020learning}. Though it leverages significantly larger number of learnable parameters and its estimations exhibit the similar image quality metric values, the U-Net produces major visual artifacts, which suggests that including the physics of the system $A$ makes the training process more efficient.

\begin{figure*}[t!]
    \centering
    \includegraphics[width=\textwidth]{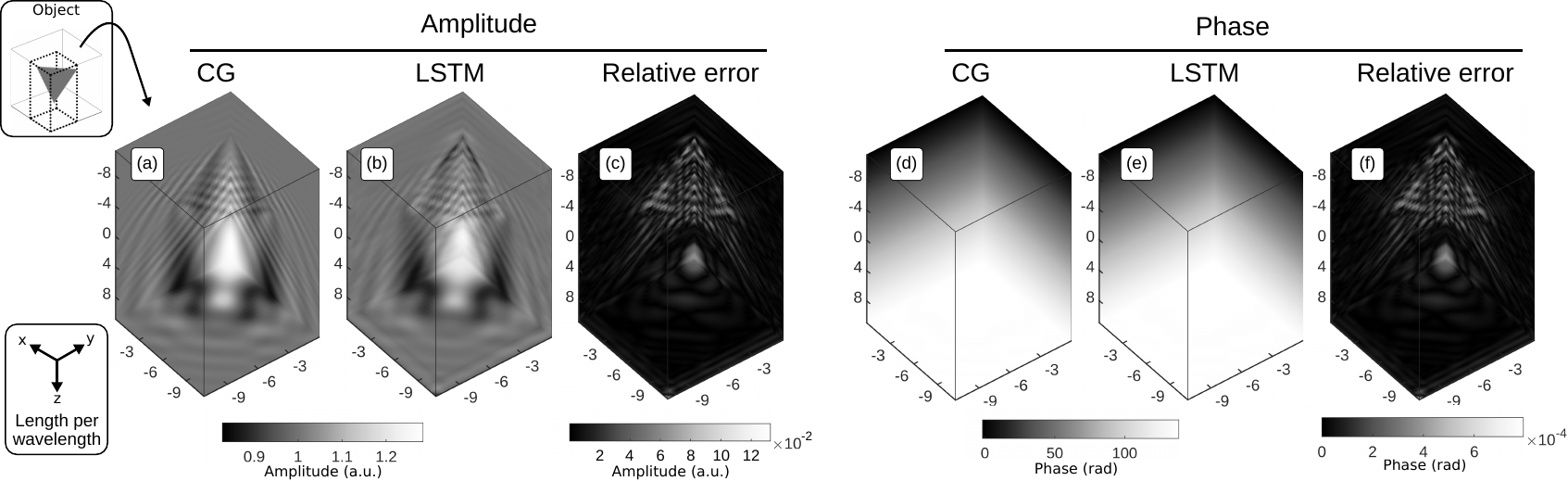}
    \caption{Comparison of the LSE with the CG and the LSTM on a tetrahedral object for the optical scattering problem. The computational box and the location of the object is drawn in the subfigure at the left top corner. The scattered field in the dotted region of this subfigure is shown here. (a),(b) The CG and the LSTM estimations on the amplitude. (c) The relative error in the amplitude. (d),(e) The CG and the LSTM estimations on the unwrapped phase. (f) The relative error in the unwrapped phase. An unit length in this figure corresponds to $\lambda$.}
    \label{fig:complex_shapes}
\end{figure*}

Interestingly, if we pass test examples with shapes completely different to the train examples, {\it e.g.} complex polyhedra, the LSTM still appears to well approximate the CG solution, as shown in Fig.~\ref{fig:complex_shapes}. The LSTM results in Fig.~\ref{fig:complex_shapes} suffer from more visual artifacts at which the incident field first contacts the tetrahedral object, compared to other regions. Such artifacts can be attributed to the difference between the surfaces of the objects in the training examples and polyhedral objects. In other words, polyhedral objects have relatively sharp edges and vertices that cannot be experienced during the training, which may be able to worsen the estimation quality. In the Supplemental Material, we test the LSTM with 5 objects of polyhedral shapes other than the tetrahedron, which also displays the similar behavior. Nonetheless, the overall estimation retains physically reasonable quality. As listed in Table~\ref{tab:cmp_shp}, the image quality metrics for such polyhedral objects also show tolerable values. The overall agreement between the results from the CG and the LSTM indirectly indicates that the physical information from $A$ is properly fused into the training process and the LSTM layer acts as an actual proximal operator to increase the convergence of the iterative algorithm. Otherwise, \textit{e.g.} if the LSTM is merely overfitted to the spherical training examples, the LSTM would map the output of the physics layer to a wrong space, leading to a poor convergence.

\textit{Numerical study on Coulomb scattering.}---The validity of the CG method on Coulomb potential is tested with an analytic solution. For this test, we exclude the screening and the singularity regularization, {\it i.e.} $V(\vcr{r}) = 2/|\vcr{r}|$ is replaced by (\ref{eq:LSCoulomb}). Detailed information on the computational architecture and the result is presented in the Supplemental Material. The comparison shows that the LSE solved with the CG well approximates the interference near the center of the electron beam and the numerical difference to the analytical solution is tolerable, implying that the CG may be applicable to estimate Coulomb scattering.

\begin{figure*}[t!]
    \centering
    \includegraphics[width=\textwidth]{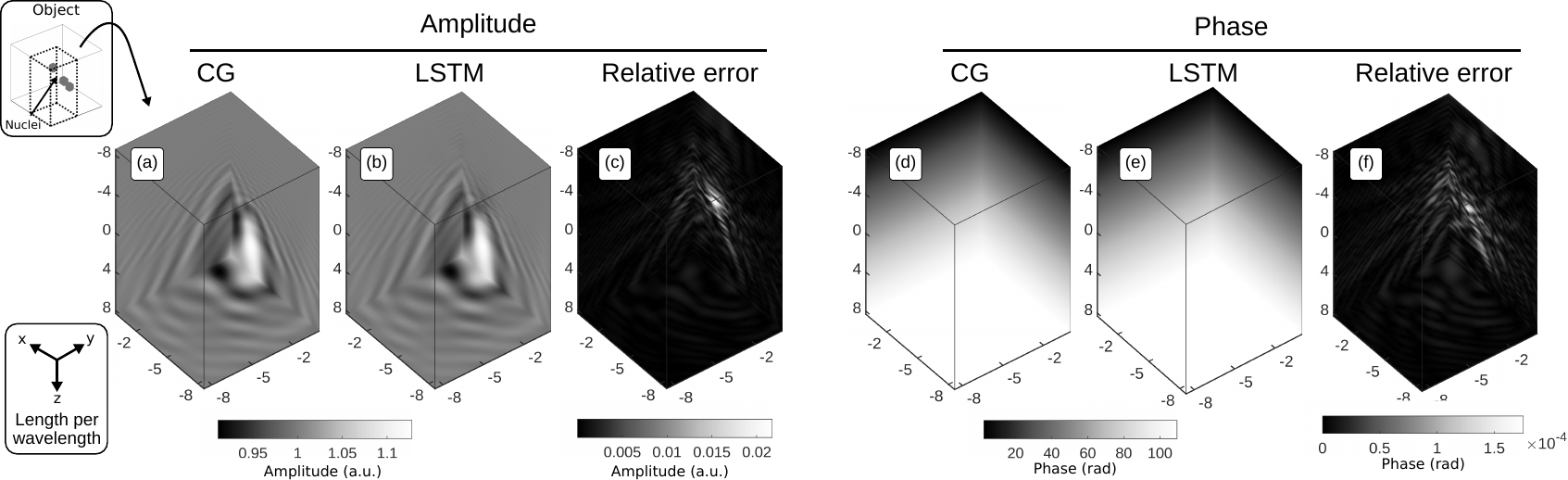}
    \caption{Comparison of the LSE with the CG and the LSTM on a test object consisting of three nuclei for the quantum scattering problem. The computational box and the location of the object is drawn in the subfigure at the left top corner. The scattered field in the dotted region of this subfigure is shown here. (a),(b) The CG and the LSTM estimations on the amplitude. (c) The relative error in the amplitude. (d),(e) The CG and the LSTM estimations on the unwrapped phase. (f) The relative error in the unwrapped phase. An unit length in this figure corresponds to $\lambda$.}
    \label{fig:CGvsLSTM_quantum}
\end{figure*}

Coulomb scattering expected by the LSTM is shown in Fig.~\ref{fig:CGvsLSTM_quantum}. As in the optical scattering case, visually the LSTM approximates the CG solutions with small deviation. The region that exhibits somewhat high relative error is in the vicinity of the two nuclei where the potentials are highly overlapping. Since the potential profile becomes deep and its shape complex, this implies that it is difficult for the LSTM to learn the regularizer. Table~\ref{tab:quantum_lstm} outlines the numerical assessment of the LSTM performance. Among the 300 test samples, the relative error per voxel is comparable to $1.0\times 10^{-3}$, which, combined with the high SSIM and the PSNR values, means that the LSTM can be utilized to estimate the solution of the LSE not only for the optical potentials but the quantum mechanical potentials.

\begin{table}[b!]
\caption{\label{tab:quantum_lstm} The image quality assessment of the LSTM on the 300 test examples for Coulomb scattering problem. The scattered fields estimated by solving the LSE with the CG are set as the ground truth images. Relative error in this table means the relative error per voxel.}
\begin{ruledtabular}
\begin{tabular}{cccc} % left, center, decimal, right
\textrm{Component}&
{\textrm{Relative error}}&
{\textrm{SSIM}}&
{\textrm{PSNR}}\\
\colrule
\textrm{Amplitude} & 0.001 & 0.995 & 58.297\\
\textrm{Phase} & $4.2\times 10^{-4}$ & 0.999 & 35.788\\
\textrm{Unwrapped phase} & $7.3\times 10^{-6}$ & 1.000 & 98.777\\
\end{tabular}
\end{ruledtabular}
\end{table}

\textit{Conclusion.}---Numerous physical systems have been formulated as the linear inverse problem and there have been corresponding studies to solve Eq.~(\ref{eq:Ax=b}), including not only classical optimization methods but data-driven approaches and the neural networks \cite{tang2017study, rudy2017data, raissi2019physics}. Compared to the previous studies, we tailor the performance of the neural networks by explicitly informing the physics of the system $A$. Furthermore, our LSTM generalizes well to unseen examples by explicitly learn the regularizating prior underlying the specific scattering matrix $A$. Hence, it can be expected that our work can contribute to solving physical problems other than the LSE.

Recently, an alternative to LSTM known as Gated Recurrent Unit (GRU) was proposed \cite{kang2020limited} and shown to generalize better in certain machine vision and robotic manipulation problems. Even though we did not investigate GRU in this study, it certainly shows merit for future investigations \cite{kang2020limited}.

\begin{acknowledgments}
This work is supported by the Intelligence Advanced Research Projects Activity (IARPA) No. FA8650-17-C-9113 and Singapore-MIT Alliance for Research and Technology (SMART) Grant No. 015824-00169.
\end{acknowledgments}

%https://www.physics.uci.edu/~silverma/revtex4.html
\providecommand{\noopsort}[1]{}\providecommand{\singleletter}[1]{#1}%


\begin{thebibliography}{28}%
\makeatletter
\providecommand \@ifxundefined [1]{%
 \@ifx{#1\undefined}
}%
\providecommand \@ifnum [1]{%
 \ifnum #1\expandafter \@firstoftwo
 \else \expandafter \@secondoftwo
 \fi
}%
\providecommand \@ifx [1]{%
 \ifx #1\expandafter \@firstoftwo
 \else \expandafter \@secondoftwo
 \fi
}%
\providecommand \natexlab [1]{#1}%
\providecommand \enquote  [1]{``#1''}%
\providecommand \bibnamefont  [1]{#1}%
\providecommand \bibfnamefont [1]{#1}%
\providecommand \citenamefont [1]{#1}%
\providecommand \href@noop [0]{\@secondoftwo}%
\providecommand \href [0]{\begingroup \@sanitize@url \@href}%
\providecommand \@href[1]{\@@startlink{#1}\@@href}%
\providecommand \@@href[1]{\endgroup#1\@@endlink}%
\providecommand \@sanitize@url [0]{\catcode `\\12\catcode `\$12\catcode
  `\&12\catcode `\#12\catcode `\^12\catcode `\_12\catcode `\%12\relax}%
\providecommand \@@startlink[1]{}%
\providecommand \@@endlink[0]{}%
\providecommand \url  [0]{\begingroup\@sanitize@url \@url }%
\providecommand \@url [1]{\endgroup\@href {#1}{\urlprefix }}%
\providecommand \urlprefix  [0]{URL }%
\providecommand \Eprint [0]{\href }%
\providecommand \doibase [0]{http://dx.doi.org/}%
\providecommand \selectlanguage [0]{\@gobble}%
\providecommand \bibinfo  [0]{\@secondoftwo}%
\providecommand \bibfield  [0]{\@secondoftwo}%
\providecommand \translation [1]{[#1]}%
\providecommand \BibitemOpen [0]{}%
\providecommand \bibitemStop [0]{}%
\providecommand \bibitemNoStop [0]{.\EOS\space}%
\providecommand \EOS [0]{\spacefactor3000\relax}%
\providecommand \BibitemShut  [1]{\csname bibitem#1\endcsname}%
\let\auto@bib@innerbib\@empty
%</preamble>
\bibitem [{\citenamefont {Vainikko}(2000)}]{vainikko2000fast}%
  \BibitemOpen
  \bibfield  {author} {\bibinfo {author} {\bibfnamefont {G.}~\bibnamefont
  {Vainikko}},\ }in\ \href@noop {} {\emph {\bibinfo {booktitle} {Direct and
  inverse problems of mathematical physics}}}\ (\bibinfo  {publisher}
  {Springer},\ \bibinfo {year} {2000})\ pp.\ \bibinfo {pages}
  {423--440}\BibitemShut {NoStop}%
\bibitem [{\citenamefont {Zepeda-N{\'u}{\~n}ez}\ and\ \citenamefont
  {Zhao}(2016)}]{zepeda2016fast}%
  \BibitemOpen
  \bibfield  {author} {\bibinfo {author} {\bibfnamefont {L.}~\bibnamefont
  {Zepeda-N{\'u}{\~n}ez}}\ and\ \bibinfo {author} {\bibfnamefont
  {H.}~\bibnamefont {Zhao}},\ }\href@noop {} {\bibfield  {journal} {\bibinfo
  {journal} {SIAM Journal on Scientific Computing}\ }\textbf {\bibinfo {volume}
  {38}},\ \bibinfo {pages} {B866} (\bibinfo {year} {2016})}\BibitemShut
  {NoStop}%
\bibitem [{\citenamefont {Ying}(2015)}]{ying2015sparsifying}%
  \BibitemOpen
  \bibfield  {author} {\bibinfo {author} {\bibfnamefont {L.}~\bibnamefont
  {Ying}},\ }\href@noop {} {\bibfield  {journal} {\bibinfo  {journal}
  {Multiscale Modeling \& Simulation}\ }\textbf {\bibinfo {volume} {13}},\
  \bibinfo {pages} {644} (\bibinfo {year} {2015})}\BibitemShut {NoStop}%
\bibitem [{\citenamefont {Gockenbach}(2005)}]{gockenbach2005partial}%
  \BibitemOpen
  \bibfield  {author} {\bibinfo {author} {\bibfnamefont {M.~S.}\ \bibnamefont
  {Gockenbach}},\ }\href@noop {} {\emph {\bibinfo {title} {Partial differential
  equations: analytical and numerical methods}}},\ Vol.\ \bibinfo {volume}
  {122}\ (\bibinfo  {publisher} {Siam},\ \bibinfo {year} {2005})\BibitemShut
  {NoStop}%
\bibitem [{\citenamefont {{\^S}ol{\'\i}n}(2005)}]{solin2005partial}%
  \BibitemOpen
  \bibfield  {author} {\bibinfo {author} {\bibfnamefont {P.}~\bibnamefont
  {{\^S}ol{\'\i}n}},\ }\href@noop {} {\emph {\bibinfo {title} {Partial
  differential equations and the finite element method}}},\ Vol.~\bibinfo
  {volume} {73}\ (\bibinfo  {publisher} {John Wiley \& Sons},\ \bibinfo {year}
  {2005})\BibitemShut {NoStop}%
\bibitem [{\citenamefont {Eason}(1976)}]{eason1976review}%
  \BibitemOpen
  \bibfield  {author} {\bibinfo {author} {\bibfnamefont {E.~D.}\ \bibnamefont
  {Eason}},\ }\href@noop {} {\bibfield  {journal} {\bibinfo  {journal}
  {International journal for numerical methods in engineering}\ }\textbf
  {\bibinfo {volume} {10}},\ \bibinfo {pages} {1021} (\bibinfo {year}
  {1976})}\BibitemShut {NoStop}%
\bibitem [{\citenamefont {Gould}\ \emph {et~al.}(2007)\citenamefont {Gould},
  \citenamefont {Scott},\ and\ \citenamefont {Hu}}]{gould2007numerical}%
  \BibitemOpen
  \bibfield  {author} {\bibinfo {author} {\bibfnamefont {N.~I.}\ \bibnamefont
  {Gould}}, \bibinfo {author} {\bibfnamefont {J.~A.}\ \bibnamefont {Scott}}, \
  and\ \bibinfo {author} {\bibfnamefont {Y.}~\bibnamefont {Hu}},\ }\href@noop
  {} {\bibfield  {journal} {\bibinfo  {journal} {ACM Transactions on
  Mathematical Software (TOMS)}\ }\textbf {\bibinfo {volume} {33}},\ \bibinfo
  {pages} {10} (\bibinfo {year} {2007})}\BibitemShut {NoStop}%
\bibitem [{\citenamefont {Starck}\ and\ \citenamefont
  {Fadili}(2009)}]{starck2009overview}%
  \BibitemOpen
  \bibfield  {author} {\bibinfo {author} {\bibfnamefont {J.-L.}\ \bibnamefont
  {Starck}}\ and\ \bibinfo {author} {\bibfnamefont {M.-J.}\ \bibnamefont
  {Fadili}},\ }in\ \href@noop {} {\emph {\bibinfo {booktitle} {2009 16th IEEE
  International Conference on Image Processing (ICIP)}}}\ (\bibinfo
  {organization} {IEEE},\ \bibinfo {year} {2009})\ pp.\ \bibinfo {pages}
  {1453--1456}\BibitemShut {NoStop}%
\bibitem [{\citenamefont {Mallat}(1999)}]{mallat1999wavelet}%
  \BibitemOpen
  \bibfield  {author} {\bibinfo {author} {\bibfnamefont {S.}~\bibnamefont
  {Mallat}},\ }\href@noop {} {\emph {\bibinfo {title} {A wavelet tour of signal
  processing}}}\ (\bibinfo  {publisher} {Elsevier},\ \bibinfo {year}
  {1999})\BibitemShut {NoStop}%
\bibitem [{\citenamefont {Aharon}\ \emph {et~al.}(2006)\citenamefont {Aharon},
  \citenamefont {Elad},\ and\ \citenamefont {Bruckstein}}]{aharon2006k}%
  \BibitemOpen
  \bibfield  {author} {\bibinfo {author} {\bibfnamefont {M.}~\bibnamefont
  {Aharon}}, \bibinfo {author} {\bibfnamefont {M.}~\bibnamefont {Elad}}, \ and\
  \bibinfo {author} {\bibfnamefont {A.}~\bibnamefont {Bruckstein}},\
  }\href@noop {} {\bibfield  {journal} {\bibinfo  {journal} {IEEE Transactions
  on signal processing}\ }\textbf {\bibinfo {volume} {54}},\ \bibinfo {pages}
  {4311} (\bibinfo {year} {2006})}\BibitemShut {NoStop}%
\bibitem [{\citenamefont {Elad}\ and\ \citenamefont
  {Aharon}(2006)}]{inv:elad2006image}%
  \BibitemOpen
  \bibfield  {author} {\bibinfo {author} {\bibfnamefont {M.}~\bibnamefont
  {Elad}}\ and\ \bibinfo {author} {\bibfnamefont {M.}~\bibnamefont {Aharon}},\
  }in\ \href@noop {} {\emph {\bibinfo {booktitle} {IEEE Computer Society
  Conference on Computer Vision and Pattern Recognition}}},\ Vol.~\bibinfo
  {volume} {1}\ (\bibinfo {organization} {IEEE},\ \bibinfo {year} {2006})\ pp.\
  \bibinfo {pages} {895--900}\BibitemShut {NoStop}%
\bibitem [{\citenamefont {Daubechies}\ \emph {et~al.}(2004)\citenamefont
  {Daubechies}, \citenamefont {Defrise},\ and\ \citenamefont
  {Mol}}]{inv:daubechies04}%
  \BibitemOpen
  \bibfield  {author} {\bibinfo {author} {\bibfnamefont {I.}~\bibnamefont
  {Daubechies}}, \bibinfo {author} {\bibfnamefont {M.}~\bibnamefont {Defrise}},
  \ and\ \bibinfo {author} {\bibfnamefont {C.~D.}\ \bibnamefont {Mol}},\
  }\href@noop {} {\bibfield  {journal} {\bibinfo  {journal} {Comm. Pure Appl.
  Math.}\ }\textbf {\bibinfo {volume} {57}},\ \bibinfo {pages} {1413} (\bibinfo
  {year} {2004})}\BibitemShut {NoStop}%
\bibitem [{\citenamefont {Beck}\ and\ \citenamefont
  {Teboulle}(2009)}]{beck2009fast}%
  \BibitemOpen
  \bibfield  {author} {\bibinfo {author} {\bibfnamefont {A.}~\bibnamefont
  {Beck}}\ and\ \bibinfo {author} {\bibfnamefont {M.}~\bibnamefont
  {Teboulle}},\ }\href@noop {} {\bibfield  {journal} {\bibinfo  {journal} {SIAM
  journal on imaging sciences}\ }\textbf {\bibinfo {volume} {2}},\ \bibinfo
  {pages} {183} (\bibinfo {year} {2009})}\BibitemShut {NoStop}%
\bibitem [{\citenamefont {Kamilov}\ and\ \citenamefont
  {Mansour}(2016)}]{kamilov2016learning}%
  \BibitemOpen
  \bibfield  {author} {\bibinfo {author} {\bibfnamefont {U.~S.}\ \bibnamefont
  {Kamilov}}\ and\ \bibinfo {author} {\bibfnamefont {H.}~\bibnamefont
  {Mansour}},\ }\href@noop {} {\bibfield  {journal} {\bibinfo  {journal} {IEEE
  Signal Processing Letters}\ }\textbf {\bibinfo {volume} {23}},\ \bibinfo
  {pages} {747} (\bibinfo {year} {2016})}\BibitemShut {NoStop}%
\bibitem [{\citenamefont {Mardani}\ \emph {et~al.}(2017)\citenamefont
  {Mardani}, \citenamefont {Monajemi}, \citenamefont {Papyan}, \citenamefont
  {Vasanawala}, \citenamefont {Donoho},\ and\ \citenamefont
  {Pauly}}]{mardani2017recurrent}%
  \BibitemOpen
  \bibfield  {author} {\bibinfo {author} {\bibfnamefont {M.}~\bibnamefont
  {Mardani}}, \bibinfo {author} {\bibfnamefont {H.}~\bibnamefont {Monajemi}},
  \bibinfo {author} {\bibfnamefont {V.}~\bibnamefont {Papyan}}, \bibinfo
  {author} {\bibfnamefont {S.}~\bibnamefont {Vasanawala}}, \bibinfo {author}
  {\bibfnamefont {D.}~\bibnamefont {Donoho}}, \ and\ \bibinfo {author}
  {\bibfnamefont {J.}~\bibnamefont {Pauly}},\ }\href@noop {} {\bibfield
  {journal} {\bibinfo  {journal} {arXiv preprint arXiv:1711.10046}\ } (\bibinfo
  {year} {2017})}\BibitemShut {NoStop}%
\bibitem [{\citenamefont {Hochreiter}(1991)}]{hochreiter1991untersuchungen}%
  \BibitemOpen
  \bibfield  {author} {\bibinfo {author} {\bibfnamefont {S.}~\bibnamefont
  {Hochreiter}},\ }\href@noop {} {\bibfield  {journal} {\bibinfo  {journal}
  {Diploma, Technische Universit{\"a}t M{\"u}nchen}\ }\textbf {\bibinfo
  {volume} {91}} (\bibinfo {year} {1991})}\BibitemShut {NoStop}%
\bibitem [{\citenamefont {Hochreiter}\ and\ \citenamefont
  {Schmidhuber}(1997)}]{hochreiter1997long}%
  \BibitemOpen
  \bibfield  {author} {\bibinfo {author} {\bibfnamefont {S.}~\bibnamefont
  {Hochreiter}}\ and\ \bibinfo {author} {\bibfnamefont {J.}~\bibnamefont
  {Schmidhuber}},\ }\href@noop {} {\bibfield  {journal} {\bibinfo  {journal}
  {Neural computation}\ }\textbf {\bibinfo {volume} {9}},\ \bibinfo {pages}
  {1735} (\bibinfo {year} {1997})}\BibitemShut {NoStop}%
\bibitem [{\citenamefont {Gonz{\'a}lez-Espinoza}\ \emph
  {et~al.}(2016)\citenamefont {Gonz{\'a}lez-Espinoza}, \citenamefont {Ayers},
  \citenamefont {Karwowski},\ and\ \citenamefont {Savin}}]{gonzalez2016smooth}%
  \BibitemOpen
  \bibfield  {author} {\bibinfo {author} {\bibfnamefont {C.~E.}\ \bibnamefont
  {Gonz{\'a}lez-Espinoza}}, \bibinfo {author} {\bibfnamefont {P.~W.}\
  \bibnamefont {Ayers}}, \bibinfo {author} {\bibfnamefont {J.}~\bibnamefont
  {Karwowski}}, \ and\ \bibinfo {author} {\bibfnamefont {A.}~\bibnamefont
  {Savin}},\ }\href@noop {} {\bibfield  {journal} {\bibinfo  {journal}
  {Theoretical Chemistry Accounts}\ }\textbf {\bibinfo {volume} {135}},\
  \bibinfo {pages} {256} (\bibinfo {year} {2016})}\BibitemShut {NoStop}%
\bibitem [{\citenamefont {Pham}\ \emph {et~al.}(2020)\citenamefont {Pham},
  \citenamefont {Soubies}, \citenamefont {Ayoub}, \citenamefont {Lim},
  \citenamefont {Psaltis},\ and\ \citenamefont {Unser}}]{pham2020three}%
  \BibitemOpen
  \bibfield  {author} {\bibinfo {author} {\bibfnamefont {T.-a.}\ \bibnamefont
  {Pham}}, \bibinfo {author} {\bibfnamefont {E.}~\bibnamefont {Soubies}},
  \bibinfo {author} {\bibfnamefont {A.}~\bibnamefont {Ayoub}}, \bibinfo
  {author} {\bibfnamefont {J.}~\bibnamefont {Lim}}, \bibinfo {author}
  {\bibfnamefont {D.}~\bibnamefont {Psaltis}}, \ and\ \bibinfo {author}
  {\bibfnamefont {M.}~\bibnamefont {Unser}},\ }\href@noop {} {\bibfield
  {journal} {\bibinfo  {journal} {IEEE Transactions on Computational Imaging}\
  }\textbf {\bibinfo {volume} {6}},\ \bibinfo {pages} {727} (\bibinfo {year}
  {2020})}\BibitemShut {NoStop}%
\bibitem [{\citenamefont {Goy}\ \emph {et~al.}(2019)\citenamefont {Goy},
  \citenamefont {Rughoobur}, \citenamefont {Li}, \citenamefont {Arthur},
  \citenamefont {Akinwande},\ and\ \citenamefont {Barbastathis}}]{goy2019high}%
  \BibitemOpen
  \bibfield  {author} {\bibinfo {author} {\bibfnamefont {A.}~\bibnamefont
  {Goy}}, \bibinfo {author} {\bibfnamefont {G.}~\bibnamefont {Rughoobur}},
  \bibinfo {author} {\bibfnamefont {S.}~\bibnamefont {Li}}, \bibinfo {author}
  {\bibfnamefont {K.}~\bibnamefont {Arthur}}, \bibinfo {author} {\bibfnamefont
  {A.~I.}\ \bibnamefont {Akinwande}}, \ and\ \bibinfo {author} {\bibfnamefont
  {G.}~\bibnamefont {Barbastathis}},\ }\href@noop {} {\bibfield  {journal}
  {\bibinfo  {journal} {Proceedings of the National Academy of Sciences}\
  }\textbf {\bibinfo {volume} {116}},\ \bibinfo {pages} {19848} (\bibinfo
  {year} {2019})}\BibitemShut {NoStop}%
\bibitem [{\citenamefont {Wang}\ \emph {et~al.}(2004)\citenamefont {Wang},
  \citenamefont {Bovik}, \citenamefont {Sheikh},\ and\ \citenamefont
  {Simoncelli}}]{wang2004image}%
  \BibitemOpen
  \bibfield  {author} {\bibinfo {author} {\bibfnamefont {Z.}~\bibnamefont
  {Wang}}, \bibinfo {author} {\bibfnamefont {A.~C.}\ \bibnamefont {Bovik}},
  \bibinfo {author} {\bibfnamefont {H.~R.}\ \bibnamefont {Sheikh}}, \ and\
  \bibinfo {author} {\bibfnamefont {E.~P.}\ \bibnamefont {Simoncelli}},\
  }\href@noop {} {\bibfield  {journal} {\bibinfo  {journal} {IEEE transactions
  on image processing}\ }\textbf {\bibinfo {volume} {13}},\ \bibinfo {pages}
  {600} (\bibinfo {year} {2004})}\BibitemShut {NoStop}%
\bibitem [{\citenamefont {Ronneberger}\ \emph {et~al.}(2015)\citenamefont
  {Ronneberger}, \citenamefont {Fischer},\ and\ \citenamefont
  {Brox}}]{ronneberger2015u}%
  \BibitemOpen
  \bibfield  {author} {\bibinfo {author} {\bibfnamefont {O.}~\bibnamefont
  {Ronneberger}}, \bibinfo {author} {\bibfnamefont {P.}~\bibnamefont
  {Fischer}}, \ and\ \bibinfo {author} {\bibfnamefont {T.}~\bibnamefont
  {Brox}},\ }in\ \href@noop {} {\emph {\bibinfo {booktitle} {International
  Conference on Medical image computing and computer-assisted intervention}}}\
  (\bibinfo {organization} {Springer},\ \bibinfo {year} {2015})\ pp.\ \bibinfo
  {pages} {234--241}\BibitemShut {NoStop}%
\bibitem [{\citenamefont {Guan}\ \emph {et~al.}(2019)\citenamefont {Guan},
  \citenamefont {Khan}, \citenamefont {Sikdar},\ and\ \citenamefont
  {Chitnis}}]{guan2019fully}%
  \BibitemOpen
  \bibfield  {author} {\bibinfo {author} {\bibfnamefont {S.}~\bibnamefont
  {Guan}}, \bibinfo {author} {\bibfnamefont {A.~A.}\ \bibnamefont {Khan}},
  \bibinfo {author} {\bibfnamefont {S.}~\bibnamefont {Sikdar}}, \ and\ \bibinfo
  {author} {\bibfnamefont {P.~V.}\ \bibnamefont {Chitnis}},\ }\href@noop {}
  {\bibfield  {journal} {\bibinfo  {journal} {IEEE journal of biomedical and
  health informatics}\ }\textbf {\bibinfo {volume} {24}},\ \bibinfo {pages}
  {568} (\bibinfo {year} {2019})}\BibitemShut {NoStop}%
\bibitem [{\citenamefont {Deng}\ \emph {et~al.}(2020)\citenamefont {Deng},
  \citenamefont {Li}, \citenamefont {Goy}, \citenamefont {Kang},\ and\
  \citenamefont {Barbastathis}}]{deng2020learning}%
  \BibitemOpen
  \bibfield  {author} {\bibinfo {author} {\bibfnamefont {M.}~\bibnamefont
  {Deng}}, \bibinfo {author} {\bibfnamefont {S.}~\bibnamefont {Li}}, \bibinfo
  {author} {\bibfnamefont {A.}~\bibnamefont {Goy}}, \bibinfo {author}
  {\bibfnamefont {I.}~\bibnamefont {Kang}}, \ and\ \bibinfo {author}
  {\bibfnamefont {G.}~\bibnamefont {Barbastathis}},\ }\href@noop {} {\bibfield
  {journal} {\bibinfo  {journal} {Light: Science \& Applications}\ }\textbf
  {\bibinfo {volume} {9}},\ \bibinfo {pages} {1} (\bibinfo {year}
  {2020})}\BibitemShut {NoStop}%
\bibitem [{\citenamefont {Tang}\ \emph {et~al.}(2017)\citenamefont {Tang},
  \citenamefont {Shan}, \citenamefont {Dang}, \citenamefont {Li}, \citenamefont
  {Yang}, \citenamefont {Xu},\ and\ \citenamefont {Wu}}]{tang2017study}%
  \BibitemOpen
  \bibfield  {author} {\bibinfo {author} {\bibfnamefont {W.}~\bibnamefont
  {Tang}}, \bibinfo {author} {\bibfnamefont {T.}~\bibnamefont {Shan}}, \bibinfo
  {author} {\bibfnamefont {X.}~\bibnamefont {Dang}}, \bibinfo {author}
  {\bibfnamefont {M.}~\bibnamefont {Li}}, \bibinfo {author} {\bibfnamefont
  {F.}~\bibnamefont {Yang}}, \bibinfo {author} {\bibfnamefont {S.}~\bibnamefont
  {Xu}}, \ and\ \bibinfo {author} {\bibfnamefont {J.}~\bibnamefont {Wu}},\ }in\
  \href@noop {} {\emph {\bibinfo {booktitle} {2017 IEEE Electrical Design of
  Advanced Packaging and Systems Symposium (EDAPS)}}}\ (\bibinfo {organization}
  {IEEE},\ \bibinfo {year} {2017})\ pp.\ \bibinfo {pages} {1--3}\BibitemShut
  {NoStop}%
\bibitem [{\citenamefont {Rudy}\ \emph {et~al.}(2017)\citenamefont {Rudy},
  \citenamefont {Brunton}, \citenamefont {Proctor},\ and\ \citenamefont
  {Kutz}}]{rudy2017data}%
  \BibitemOpen
  \bibfield  {author} {\bibinfo {author} {\bibfnamefont {S.~H.}\ \bibnamefont
  {Rudy}}, \bibinfo {author} {\bibfnamefont {S.~L.}\ \bibnamefont {Brunton}},
  \bibinfo {author} {\bibfnamefont {J.~L.}\ \bibnamefont {Proctor}}, \ and\
  \bibinfo {author} {\bibfnamefont {J.~N.}\ \bibnamefont {Kutz}},\ }\href@noop
  {} {\bibfield  {journal} {\bibinfo  {journal} {Science Advances}\ }\textbf
  {\bibinfo {volume} {3}},\ \bibinfo {pages} {e1602614} (\bibinfo {year}
  {2017})}\BibitemShut {NoStop}%
\bibitem [{\citenamefont {Raissi}\ \emph {et~al.}(2019)\citenamefont {Raissi},
  \citenamefont {Perdikaris},\ and\ \citenamefont
  {Karniadakis}}]{raissi2019physics}%
  \BibitemOpen
  \bibfield  {author} {\bibinfo {author} {\bibfnamefont {M.}~\bibnamefont
  {Raissi}}, \bibinfo {author} {\bibfnamefont {P.}~\bibnamefont {Perdikaris}},
  \ and\ \bibinfo {author} {\bibfnamefont {G.~E.}\ \bibnamefont
  {Karniadakis}},\ }\href@noop {} {\bibfield  {journal} {\bibinfo  {journal}
  {Journal of Computational Physics}\ }\textbf {\bibinfo {volume} {378}},\
  \bibinfo {pages} {686} (\bibinfo {year} {2019})}\BibitemShut {NoStop}%
\bibitem [{\citenamefont {Kang}\ \emph {et~al.}(2020)\citenamefont {Kang},
  \citenamefont {Goy},\ and\ \citenamefont {Barbastathis}}]{kang2020limited}%
  \BibitemOpen
  \bibfield  {author} {\bibinfo {author} {\bibfnamefont {I.}~\bibnamefont
  {Kang}}, \bibinfo {author} {\bibfnamefont {A.}~\bibnamefont {Goy}}, \ and\
  \bibinfo {author} {\bibfnamefont {G.}~\bibnamefont {Barbastathis}},\
  }\href@noop {} {\bibfield  {journal} {\bibinfo  {journal} {arXiv preprint
  arXiv:2007.10734}\ } (\bibinfo {year} {2020})}\BibitemShut {NoStop}%
\end{thebibliography}
\end{document}